\documentclass{iau}
\usepackage{graphicx}
\usepackage{amsmath}
\newcommand{\vb}{v_{\rm b}}
\newcommand{\vf}{v_{\rm f}}
\newcommand{\MJup}{M_{\rm Jup}}
\newcommand{\Msolar}{{\rm M_{\odot}}}   
\newcommand{\Rt}{R_{\rm t}}

\newcommand{\Mstar}{M_{\star}}
\newcommand{\Mdisc}{M_{\rm d}}
\usepackage{wrapfig}

\title[Grain growth in circumstellar discs] 
{Large grains can grow in circumstellar discs}

\author[Farzana Meru et al.]   
{\normalsize{Farzana Meru$^1$,
Marina Galvagni$^2$,
Christoph Olczak$^{3,4,5}$
 \& Pascale Garaud$^6$}}

\affiliation{$^1$Institut f\"ur Astronomie, ETH Z\"urich, Wolfgang-Pauli-Strasse 27, 8093 Z\"urich\\ email: {\tt farzana.meru@phys.ethz.ch} \\[\affilskip]
$^2$Institute of Theoretical Physics, Universit\"at Z\"urich, Winterthurerstrasse 190, 8057 Z\"urich\\
$^3$Astronomisches Rechen-Institut (ARI), Zentrum f{\"u}r Astronomie Universit{\"a}t Heidelberg, M{\"o}nchhofstrasse 12-14, 69120 Heidelberg\\
$^4$Max-Planck-Institut f{\"u}r Astronomie (MPIA), K{\"o}nigstuhl 17, 69117 Heidelberg\\
$^5$National Astronomical Observatories of China, 20A Datun Lu, Beijing 100012\\
$^6$Department of Applied Mathematics and Statistics, UCSC, 1156 High Street Santa Cruz
}

\pubyear{2013}
\volume{xxx}  
\pagerange{--}
\jname{Title of your IAU Symposium}
\editors{A.C. Editor, B.D. Editor \& C.E. Editor, eds.}
\begin{document}

\maketitle

\begin{abstract}
We perform coagulation \& fragmentation simulations to understand grain growth in T Tauri \& brown dwarf discs. We present a physically-motivated approach using a probability distribution function for the collision velocities and separating the deterministic \& stochastic velocities.  We find growth to larger sizes compared to other models. Furthermore, if brown dwarf discs are scaled-down versions of T Tauri discs (in terms of stellar \& disc mass, and disc radius), growth at the same location with respect to the outer edge occurs to similar sizes in both discs.

\keywords{methods: analytical; methods: numerical; planetary systems: protoplanetary disks}
\end{abstract}

\firstsection 
\section{Introduction}

Previous models of coagulation and fragmentation of an ensemble of dust aggregates often assumed that a collision between particles of particular sizes occurs with a single velocity.  We develop a more physically-motivated model for the collision velocities.

\section{Principles behind the new physically-motivated model}

\begin{figure}
  \begin{center}
   \includegraphics[width=0.6\columnwidth]{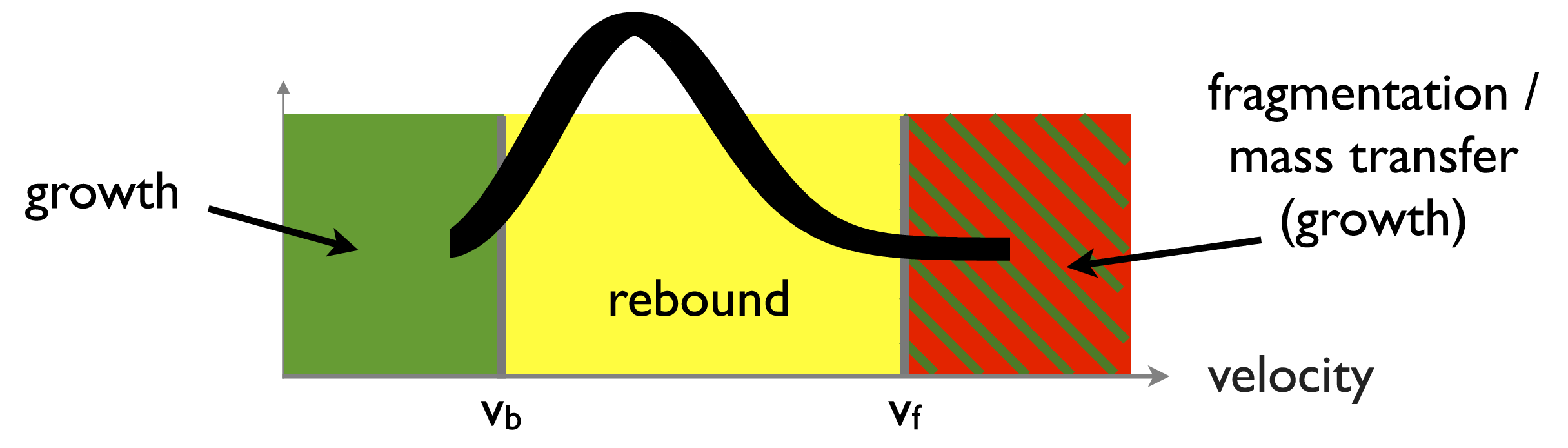}
\end{center}
\caption{Schematic diagram showing the collisional outcomes of
  sticking and bouncing at low velocities, and fragmentation and mass
  transfer at high velocities. The PDF has an expectation in one zone
  but extends into others.}
\label{fig:schematic}
\end{figure}

The key features are: (i) a particle's velocity is \emph{not} single-valued but is described by a probability distribution function (PDF); (ii) a particle's velocity in any direction is a Gaussian with a mean given by the deterministic velocity in that direction (i.e. radial drift, azimuthal drift or vertical settling) and a standard deviation given by the stochastic velocities (i.e. turbulence \& Brownian motion).  The analytical 1D PDF of \emph{relative velocities} in each direction is produced for each pair of particles, and combined to give a 3D collision velocity PDF.

We assume that collisions with velocities lower than the bouncing velocity, $\vb$, lead to growth while those higher than the fragmentation velocity, $\vf$, break apart if their mass ratio is smaller than the \emph{mass transfer} parameter and stick if it is larger.  Physically this means that collisions between unequal-sized aggregates are likely to lead to growth while equal-sized aggregates are likely to fragment - a result shown by experiments (e.g. \cite{Teiser_Wurm_highVcoll}) and simulations (\cite{Velocity_thresholds}).  At all other velocities the aggregates bounce.  Fig.~\ref{fig:schematic} shows that the final 3D PDF may cover any of these collisional outcomes - these are used to simulate the local size evolution of an ensemble of particles in a disc.

\section{Application to T Tauri and brown dwarf discs}

 \begin{figure}[h]
\begin{center}
 \includegraphics[width=0.4\columnwidth]{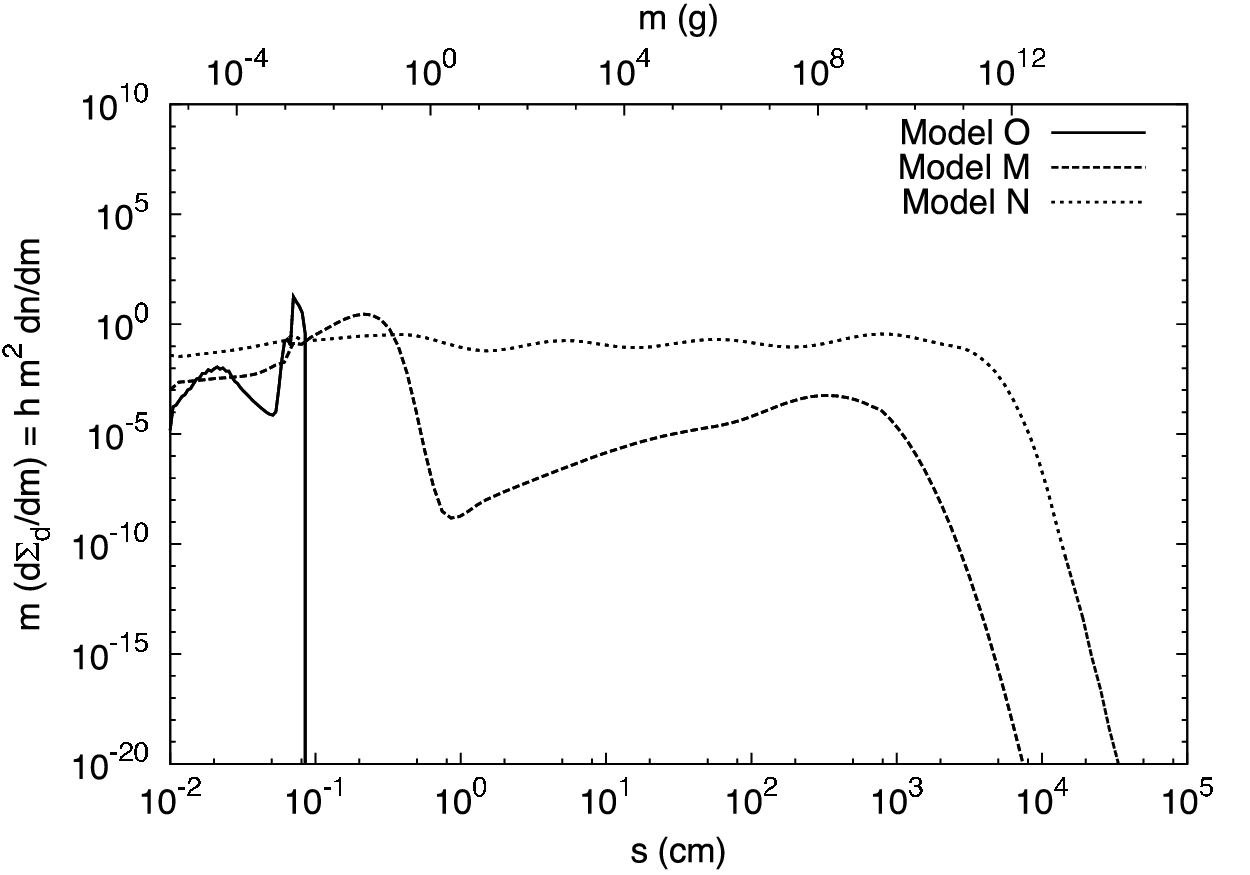} 
 \hspace{1.0cm}
   \includegraphics[width=0.3\columnwidth]{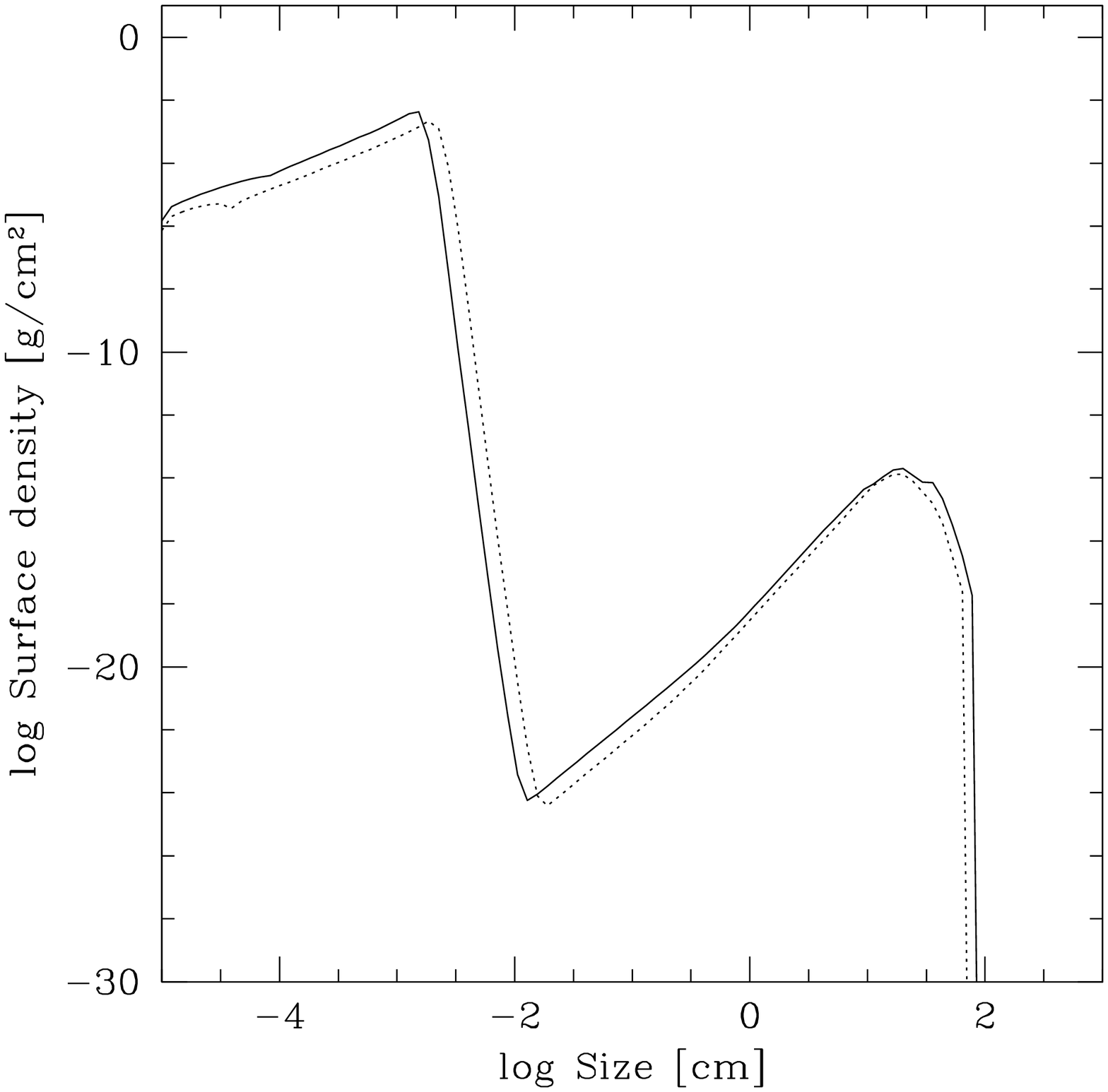}
\caption{Surface mass density distribution of particles.  Left: at 1au for a model (i) without a velocity PDF (Model O), (ii) where the stochastic velocities dominate over the deterministic ones (Model M), and (iii) the new model (Model N).  Right: for the brown dwarf disc (solid line) and T Tauri disc (dotted line) at 2/3 of the distance to the outer radius.}
   \label{fig:TT}
\end{center}
\end{figure}

We simulate grain growth locally in a disc with mass $\Mdisc = 0.05 \Mstar$ around a $0.75 \Msolar$ star (see \cite{Garaud_vel_pdf} for parameters) and find growth to larger sizes (Fig.~\ref{fig:TT}, left).  We also model growth in brown dwarf \& T Tauri discs where the latter is a scaled-up version of the former, i.e. the same disc to star mass ratio, simulated at the same location with respect to the truncation radius, $\Rt$, and where $\Mdisc$ \& $\Rt$ are set using the observed relation $\Mdisc \propto \Rt^{1.6}$ (\cite{Andrews_Md_Rt}).  We simulate growth at 10au in a brown dwarf disc with $\Mdisc = 4\times 10^{-4} \Msolar$ and $\Rt = 15 \rm au$ around a $60 \MJup$ brown dwarf, and at 60au in a T Tauri disc with $\Mdisc = 7\times 10^{-3} \Msolar$ and $\Rt = 90 \rm au$ around a $1 \Msolar$ star.  Fig.~\ref{fig:TT} (right) shows that growth at the equivalent location in both discs occurs to the same size (\cite{BD_discs_letter}).

\section{Conclusions}
We present a model for growth and fragmentation that considers a particle's velocity PDF and separates the deterministic and stochastic velocities.  We find growth to large sizes and the emergence of two particle populations.  In addition if brown dwarf discs are scaled-down versions of T Tauri discs, growth occurs to similar sizes.  Our model may potentially explain the large ($\approx$ mm-sized) grains observed in brown dwarf discs (e.g. \cite{Bouy_BDdiscs_mm}) and the long-standing problem of grain growth in discs: growing dust while maintaining a small-size population.

\end{document}